\documentclass[useAMS,usenatbib,usegraphicx]{mn2e}

\usepackage{times}
\usepackage{amsmath}

\input psfig.sty

\title[The colours of BL Lac objects]
   {The colours of BL Lac objects: a new approach to their classification}
\author[E. Massaro, R. Nesci and S. Piranomonte]
  {Enrico~Massaro$^1$,\thanks{E-mail addresses: \texttt{enrico.massaro@uniroma1.it}} 
  Roberto~Nesci$^{1}$, 
  Silvia~Piranomonte$^{2}$ \\
$^1$ Department of Physics, University of Rome ``La Sapienza'', Rome, Italy \\
$^2$ Osservatorio Astronomico di Roma, Monte Porzio Catone (RM), Italy \\
}
      
\date{Accepted 20 Feb, 2012}

\pagerange{\pageref{firstpage}--\pageref{lastpage}} \pubyear{2012}

\begin{document}
\label{firstpage}

\maketitle

\begin{abstract}
We selected a sample of 437 BL Lac objects, taken from the RomaBZCat catalogue, for 
which spectroscopic information and SDSS photometry is available. 
We propose a new classification of BL Lacs in which the sources' type is not defined 
only on the basis of the peak frequency of the synchrotron component in their
Spectral Energy Distribution (types L and H), but also on the relevance of this component
with respect to the brightness of the host galaxy (types N and G, for nuclear or galaxy
dominated sources).
We found that the SDSS colour index $u-r$=1.4 is a good separator between these two types.
We used multiband colour-colour plots to study the properties of the BL Lac classes and found
that in the X-ray to radio flux ratio vs $u-r$ plot most of the N (blue) sources are located 
in a rather narrow strip, while the G-sources (red) are spread in a large area, and most
of them are located in galaxy clusters or interacting systems, suggesting that their X-ray 
emission is not from a genuine BL Lac nucleus but it is related to their environment. 
Of the about 135 sources detected in the $\gamma$ rays by Fermi-GST, nearly all 
belong to the N-type, indicating that only this type of sources should be considered as genuine 
BL Lac nuclei.
The $J-H$, $H-K$ plot of sources detected in the 2MASS catalogue is consistent with that 
of the "bona fide" BL Lac objects, independently of their N or G classification from 
the optical indices, indicating the existence in G-type sources of a K-band excess 
possibly due to a steep, low frequency peaked emission which deserves further investigations.
We propose to use these colour plots as a further tool for searching candidate counterparts of 
newly discovered high-energy sources.

\end{abstract}
	

\section{Introduction}

The definition of BL Lac objects is somewhat ambiguous and changed depending on the 
selection criteria adopted in the surveys in which they were found.
According to the early review papers by Stein, O’Dell \& Strittmatter (1976) and 
Angel \& Stockman (1980), BL Lacs are Active Galactic Nuclei highly variable and linearly 
polarised in the radio band and with featureless optical spectra.
When emission lines were detected in a few BL Lac sources, the definition was 
modified to include the presence of these spectral features, but their restframe 
EW was limited to be lower than 5 \AA~ (Stickel et al. 1991, Stocke et al. 1991).
Another change of the BL Lac definition was introduced after the discovery in the Einstein 
Medium Sensitivity Survey (EMSS) of the so-called X-ray selected BL Lacs, which 
usually exhibit a low radio brightness (Stocke et al. 1991).
Optical counterparts of this type of BL Lacs had optical spectra either dominated
by a strong blue continuum or exhibiting a typical elliptical galaxy spectrum,
but with a low Ca H\&K break contrast.
This is defined by the ratio $C = (F_- - F_+)/F_+$, where 
$F_-$ and $F_+$ are the mean flux densities measured in small ranges (200 \AA) at 
wavelengths just lower and higher than that of the break, respectively.
Stocke et al. (1991) introduced this parameter for BL Lac selection and assumed the
upper threshold value of 0.25, quite lower than that observed in normal elliptical 
galaxies ($\sim$0.5), combined with the absence of emission lines having EW$>$ 5 \AA.
It should be also able to discriminate BL Lacs from normal galaxies exhibiting a
significant X-ray emission of non-nuclear origin, as, for example, close or inside a 
surrounding cluster. 
The threshold value of the Ca H\&K contrast was after increased to 0.4 by March\~a et.
al. (1996). 
Landt, Padovani \& Giommi (2002) proposed that the large differences in the Ca H\&K 
contrast are essentially due to the jet orientation to the line of sight and that
sources having 0.25$< C <$0.4 should be considered as a bridge between BL Lacs
and FR I radio galaxies.

Some recent surveys, aimed to the identification of new BL Lac objects or candidates,
are essentially based on the application of these criteria to a large population of
extragalactic sources and provided lists of several hundredths of objects.
The ROXA survey (Turriziani et al. 2007), for instance, considered sources detected
in all the radio, optical and X-ray bands, while Plotkin et al. (2008, 2010) searched
BL Lac objects using SDSS spectral data combined with the accurate location of radio 
sources from the FIRST radio survey (White et al. 1997).
A direct application of these criteria, without an accurate analysis of the sources'
properties and environment, could drive to consider as a genuine BL Lac object any 
AGN exhibiting a weak non-thermal emission, likely originating from the nucleus 
together with either a complete absence of emission lines or a possible occurrence 
with low EWs.
As a consequence, considering also the possibility to have a spurious association due 
to large error boxes, as those typical at high energies, the knowledge of the BL Lac 
properties, luminosity functions and cosmological evolution may be rather uncertain.
This problem becomes more relevant when associations are obtained by means of
automatic procedures, often based on non-homogenous and incomplete databases.
It is important, therefore, to perform an  extended comparative analysis of the results 
of recent surveys to establish safe criteria for defining the BL Lac characteristics.

In this paper we study a sample of BL Lacs extracted from the Roma-BZCAT catalogue
of blazars (Massaro et al. 2009, 2011a) and analyse their population properties 
using a direct approach based on the flux ratios in different frequency ranges.
We are able to derive some criteria which improve the efficiency in discriminating
genuine BL Lacs from other AGNs.
These criteria can be also applied to gain confidence in the search of possible
counterparts of high-energy sources, whose large error boxes can provide a high
probability for spurious associations.

\begin{figure}
\vspace{0.0cm}
\centering
\includegraphics[height=8.4cm,angle=-90]{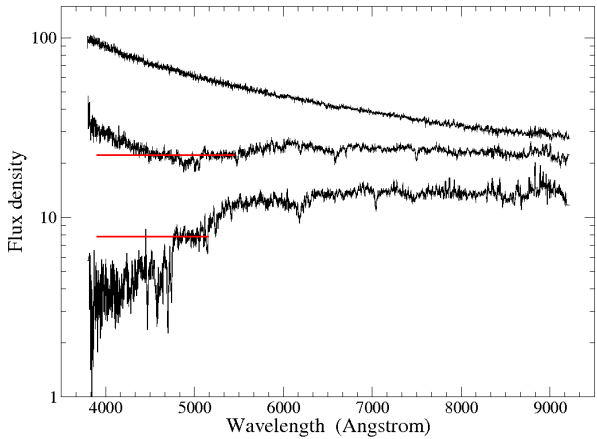}
\caption{SDSS spectra of three BL Lac objects showing different ratio of
the nuclear to galaxian luminosities. 
From top to bottom: BZB J1120+4212 (($u-r)_{obs}$=0.4), BZB J0909+3105 ($z=0.272$, 
$(u-r)_{obs}$=1.3), and BZB J0751+2913 ($z=0.194$, $(u-r)_{obs}$=2.9).
Red lines, corresponding to the mean flux level in a 400 \AA wide wavelength interval
redwards the Ca H\&K break, are used to distinguish N and G type BL Lac objects.
}
\label{fig1}
\end{figure}

\section{Selection of BL Lac samples}

The master sample of BL Lac objects used in our classification study 
was extracted from the 3rd edition of the Roma-BZCAT
\footnote{The Roma-BZCAT is available at the web site http://asdc.asi.it/bzcat}.
This catalogue is a compilation of literature data on blazars and candidates 
mainly aimed to the search of possible counterparts of high energy sources.
It provides a rich list of BL Lac objects, selected according
to the criteria currently adopted in the literature.
BL Lac objects are divided into two groups on the basis of the 
availability of their optical spectra (or of a detailed description).
Sources claimed in the literature as BL Lacs, without a published spectrum 
or a good spectral information are classified as ``candidates''.
These candidates, therefore, are not included in the sample because spectra are 
necessary to evaluate the relevance between the nuclear and galaxian components. 

Multiband photometric data are also useful for evaluating the relevance of the nuclear 
emission with respect to the host galaxy, particularly at frequencies where the galaxian 
contribution is much different from the nuclear one.
An homogeneous collection of photometric data, including the $u$ band, is available
only for sources in the region of the sky covered by the Sloan Digital Sky Survey
(SDSS, for the DR7 version see Abazajian et al. 2009), while other photometric data in the 
Johnson-Morgan $U$ band are reported in the literature only for a rather small number of 
bright and well studied objects.
We decided, therefore, to include in our master sample only those BL Lac objects with
available SDSS photometric data.
The two well known and very close BL Lac objects Mrk 421 and Mrk 501 were excluded from
the sample because the apparent size of their galaxies is too large and aperture effect on 
the photometric data would modify the colours.
We thus obtained a number of 683 sources over a total number of 937 catalogued BL Lacs.

Unfortunately, not all the sources included in the master sample (hereafter named B0) 
have been observed in all the frequency ranges, and therefore we had to define a number 
of subsamples useful for more homogeneous population studies.
Therefore, we defined three subsamples including sources detected in the near IR, and 
X-ray, and $\gamma$-ray band.
The Roma-BZCAT lists the X-ray flux in the band 0.1--2.4 keV, mainly from RASS but other
observatories like XMM and Swift are also considered, particularly for recently discovered
sources.
$\gamma$-ray fluxes were obtained from the 1FGL catalogue (Abdo et al. 2010a), and
near IR data from the 2MASS database (Cutri et al. 2003).
The three subsamples are named B1-X, B1-G, and B1-IR and the numbers of sources are 423, 
109, and 295, respectively.
Radio flux at 1.4 GHz is available for all the sources, because radio detection is requested 
to be included in the Roma-BZCAT.

\section{BL Lac colours and classification} 

\subsection{L-H classification}

Multifrequency observations have shown that the Spectral Energy 
Distribution (SED) of BL Lac objects is characterised by two broad features. 
The feature peaking at lower energy is generally explained in terms of 
synchrotron emission from relativistic electrons moving down a jet pointing 
close to the observers' line of sight whereas the second feature, peaking 
at higher energies, is likely due to the Inverse Compton upscattering of 
low energy photons by the same electron population.
A first, now widely used, physical classification of BL Lac objects based on 
the frequency of the synchrotron peak in the SED $\nu_p$, was introduced by Padovani 
\& Giommi (1995), 
who divided these sources in Low-frequency peaked (LBL) and High-frequency peaked 
(HBL) BL Lacs.
Recently, Abdo et al. (2010b) extended this classification to all the blazars and 
defined Low Synchrotron Peaked (LSP) blazars sources as those having $\nu_p$ lower 
than $\sim$10$^{14}$ Hz, High Synchrotron Peaked (HSP) blazars those with $\nu_p$
 higher than $\sim$10$^{15}$ Hz, while sources 
within these two limits are defined as ISP or Intermediate Synchrotron Peaked blazars 
(see also Kock et al. 1996, Laurent-Muehleisen et al. 1999, Bondi et al. 2001).
An accurate classification would require the analysis of the SED and a good estimate 
of the peak frequency of the synchrotron emission.
This analysis can be performed only for a rather small number of well studied sources 
whose SEDs have many data points covering a broad frequency range. 
For the majority of sources data are too few and not simultaneous, implying that 
results are much more uncertain.
Nieppola, Tornikoski \& Valtaoja (2006) presented a first analysis of the peak 
frequencies of a large sample of BL Lac objects including more than 300 sources.
They applied a single parabolic fit to a selection of data for each source and
estimated the peak frequency, which was used for a classification in the three types
given above, but with different frequency ranges. More specifically,
they defined Intermediate BL objects those having a synchrotron peak frequency 
between $\sim$10$^{14.5}$ Hz and $\sim$10$^{16.5}$ Hz.

\begin{figure}
\vspace{0.0cm}
\centering
\includegraphics[height=8.4cm,angle=-90]{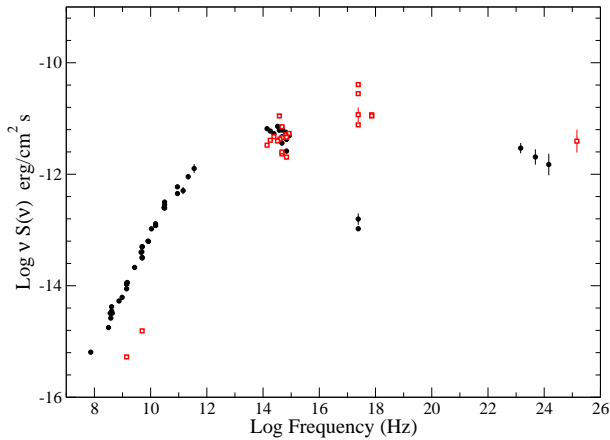}
\caption{The Spectral Energy Distributions of the BL Lac objects BZB J1150+2417 (L type, 
filled black circles) and BZB J1517+6525 (H type, red open squares) to illustrate as 
the ratio $\Phi_{XR}$ can be a good tool for this classification. 
}
\label{fig2}
\end{figure}

The sample considered by us is richer than the one of Nieppola, Tornikoski \& Valtaoja 
(2006), but common sources are only 169.
For the majority of our objects the available data are too few to try a best fit of 
the SED.
We needed, therefore, to apply a much simpler strategy based on a flux ratio between 
two largely different frequency interval.  
This simple approach does not allow us to distinguish clearly Intermediate BL Lacs 
from the other two types, and therefore throughout this paper, we will consider 
only the two LBL and HBL classes and refer shortly to them as L and H, respectively.

A useful quantity for discriminating L and H BL~Lac objects is the ratio between 
the X-ray and the radio flux.
This ratio was early used in the Sedentary Survey (Giommi et al. 2005) to select HBL 
objects and it was found to work well also for wider BL Lac samples.
The Roma-BZCAT (Massaro et al. 2011a) gives the X-ray flux in the 0.1--2.4 keV band in units 
of 10$^{-12}$ erg cm$^{-2}$ s$^{-1}$, while the radio flux density at 1.4 GHz is in mJy.
Following Maselli et al. (2010a), to express this ratio in a convenient adimensional number, 
we multiplied the radio flux density by a bandwidth $\Delta \nu$, assumed equal to 1 GHz, 
and the value of the ratio was opportunely multiplied by a scale factor; for the Roma-BZCAT
data we have:
\begin{equation}
\Phi_{XR} = 10^{2} \frac{F_X}{S_{1.4}~\Delta\nu}
\end{equation}
\noindent
In this way, the value of $F_X/S_{1.4}$ = 10$^{-11}$ erg cm$^{-2}$ s$^{-1}$ Jy$^{-1}$ 
corresponds to $\Phi_{XR}$ = 1.
We found that for well established L sources the values of $\Phi_{XR}$ are smaller than 
0.1, while for the H objects are higher than 1. 
The classification of sources having $\Phi_{XR}$ between these two values is rather 
uncertain: some of them could be more properly classified as Intermediate BL Lac objects, 
whereas for other sources this ratio could be due to brightness variations affecting 
observations with a long temporal distance, and for some others to the presence of 
additional emission components leading to a misclassification.
An example to illustrate how the ratio $\Phi_{XR}$ can discriminate between L and H objects
is given in Fig. 2, where the SEDs of the two $\gamma$-ray loud BL Lac objects BZB J1150+2417 
(L type) and BZB J1517+6525 (H type) are compared. 
The large difference between the frequencies of their synchrotron peaks (one in the infrared 
and the other in the X-rays) is very well evident and the corresponding values of $\Phi_{XR}$ 
are 0.038 for the former source and 33.7 for the latter one.
The histogram of $\Phi_{XR}$ is shown in Fig. 3: the large majority of sources in our sample
are of H type as apparent from the large bump between 1 and 10; the fraction of L sources is
rather small, despite they were the only BL Lacs discovered before the first X-ray surveys
in the '80s. 
We also compared our results with the classes given by Nieppola, Tornikoski \& Valtaoja 
(2006) and found always an excellent agreement.
Only for a quite small number of BL Lacs our types were different, but these discrepancies
were generally due to the fact that we used X-ray data published in the last few years. 

\begin{figure}
\vspace{0.0 cm}
\centering
\includegraphics[height=8.4cm,angle=-90]{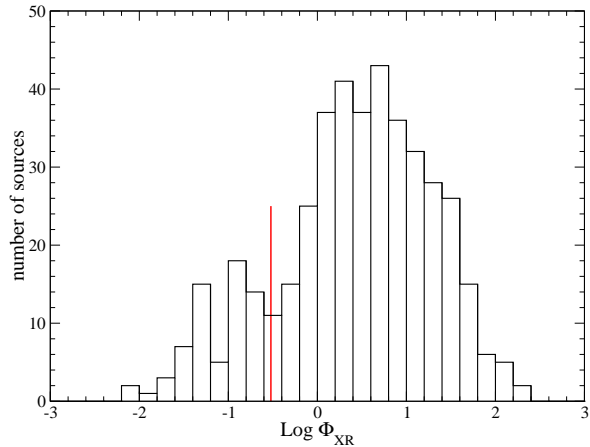}
\caption[]{Histogram of $\Phi_{XR}$.
The red line is the formal boundary adopted between L and H type.
}
\end{figure}

There are, however, a couple of caveats to be taken into account in using this parameter 
for discriminating HBL from LBL objects.
The first one is that measures of the X-ray flux are not available in the literature or 
databases for about 1/3 of the sources in the B0 sample.
This lack of data does not mean that the X-ray emission from these source is fainter 
than for other objects in the sample.
For several sources data were obtained in pointed observations of various satellites, while 
for many others the flux was derived from RASS (Brinkmann et al. 1997, Voges et al. 1999).
Considering that the flux limit of this survey is not uniform in the sky, a number of
sources could not be detected with a significance level to be included in their catalogue. 
The second caveat is that the reported X-ray flux may be occasionally contaminated either 
by a close nearby source or from emission originating in the surroundings, as for 
objects located in small clusters or groups of galaxies, or in interacting galaxy systems.
Foreground or background sources can also produce an additional spurious contribution
to the X-ray flux assigned to the BL Lac object.
These two caveats are particularly relevant for the search of BL Lac objects performed working
on databases, because the X-ray emission from many small clusters or groups is generally
not clearly identified and reported in the main catalogues.
We will discuss in detail this subject because we found that it is the reason for 
some misclassifications.

\begin{figure}
\vspace{0.0 cm}
\centering
\includegraphics[height=8.4cm,angle=-90]{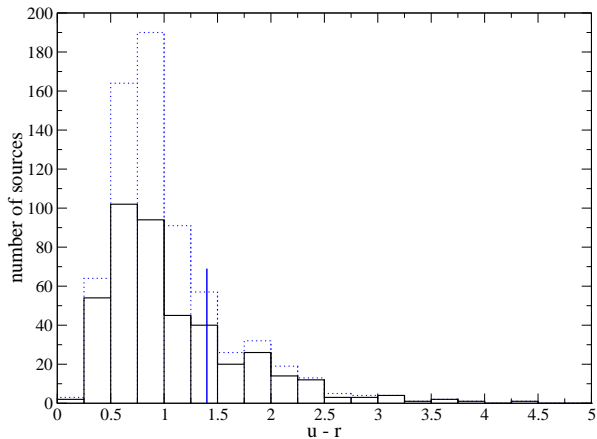}
\caption[]{Histograms of the $u - r$ colour distributions for the BL Lacs in the B1-X sample
(solid black blue histogram) and for all sources of the B0 sample (dotted blue histogram).
The vertical line divides N from G objects.
Note that the large majority of sources without a measured X-ray flux, has colours typical
of an N type BL Lac object.}
\end{figure}

\subsection{N-G classification}

Another classification of BL Lac objects can be based on the relative strength of the 
nuclear emission to the host galaxy luminosity inferred from the optical spectra.
Following the general approach that the Ca H\&K break is a good indicator of the
relevance of these two components, we considered the two following types:

-- N (or nucleus dominated) BL Lacs, for sources exhibiting a strong blue continuum
either without evidence of the Ca H\&K break or with a mean flux density at frequencies
higher than that of the Ca H\&K break having a level similar or higher than that measured 
at lower frequencies;  

-- G (or galaxy dominated) BL Lacs, for sources in which the Ca H\&K break was always 
detected and with a flux density at higher frequencies systematically lower than that
measured at lower frequencies.

The quantitative criterion for this classification is to compute the mean flux level 
in a 400 \AA~ wide wavelength interval redwards the Ca H\&K break and to verify if the 
flux at shorter wavelenghts is or not systematically lower than this value.
In Fig. 1 red lines mark the fiducial level: the first two spectra are of N type BL Lacs, 
whereas the third one is clearly of G type.
Note that in applying criterion sources with $C$ values lower than 0.4 can be of G type.


\begin{figure*}
\vspace{0.0 cm}
\centering
\includegraphics[height=13.5cm,angle=-90]{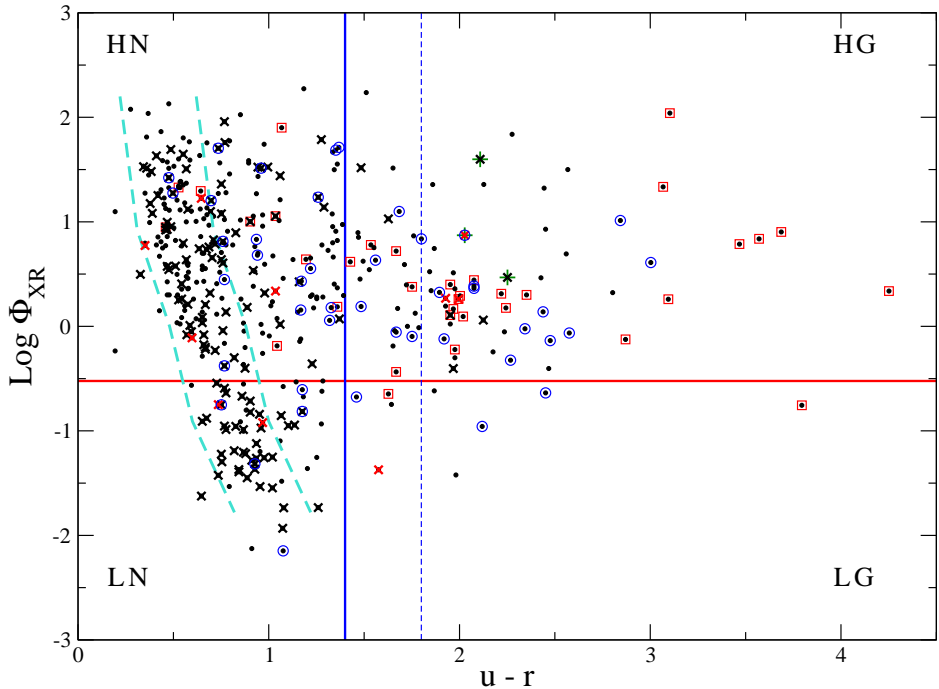}
\caption[]{Scatter plot of $Log~ \Phi_{XR}$ vs $u - r$. 
Thick dashed lines limit the high density region.
Thick small ${\bf \times}$ mark the sources detected in the $\gamma$-ray range and reported in
the 2FGL catalogue (black) or only in the 1FGL (red); large open red squares correspond to 
sources close to or in clusters; 
large open blue circles correspond to sources in small groups according to our analysis.
}
\end{figure*}

The wavelength range of spectral data is frequently limited at about 4000 \AA~ as in the 
SDSS, and therefore the level of the nuclear emission at shorter wavelengths cannot be 
well evaluated.
To extend the band and to make more quantitative the classification, we considered  
the $u - r$ colour index of the SDSS photometric system (AB magnitudes) that was 
found a simple and efficient tool for discriminating between BL Lacs of N and G type. 
The effective wavelengths of the $u$ and $r$ bandpasses are 3500 and 6250 \AA, respectively.
Considering that the minimum redshift that a BL Lac should have to move the Ca H\&K 
break up to the observer $r$ bandpass is $z\simeq$0.5 and that redshifts of BL Lacs 
are typically lower than this value, the Ca H\&K break lies between these two wavelengths 
and, therefore, the $u - r$ is a well suited parameter to measure the relevance of 
the nuclear emission with respect to that of the galaxy.
In this analysis we considered SDSS model magnitudes to take properly into account the
host galaxy contribution.

To evaluate the intrinsic $u - r$ colour we corrected the SDSS magnitudes using the $A_r$ 
absorption given in the SDSS and the extinction curve by Schlegel, Finkbeiner \& Davis (1998).
The resulting correction was computed by means of the following formula:
\begin{equation}
(u - r) = (u - r)_{obs} - 0.81~ A_r
\end{equation}
\noindent
where $A_r$ is the extinction in the $r$ band, also reported in the SDSS database.
Photometric uncertainties in the SDSS $u$ band for the large majority of objects in our 
sample are of the order of a few hundredths of magnitude and generally smaller than one 
tenth. 
For a few cases, typically very faint sources or having a remarkably high colour indices,
large errors up a $\sim$1 magnitude and even more, are given.
Considering that the resulting uncertainty on $u - r$ prevents a safe classification we
discarded the sources with a photometric error higher than 0.4.
This choice would not affect our results because these sources are practically
all of G type and their colours are so high that large errors do not modify the
classification. 
The resulting sample contained 679 objects.
Fig. 4 shows the histograms of the $u -r$ colour, after the reddening correction, for all 
the BL Lac objects in the B0 sample and for those in the B1-X. 
These two distributions are similar and the only difference is that the B0 sample contains a number
of blue objects higher than B1-X.
It is possible to calculate how large the colour $u - r$ must be to discriminate between N and 
G types.
The request that the mean flux in the $u$ band would be comparable to the extrapolation of the
level measured between 3600 \AA~ and 4000 \AA~ corresponds to  $u - r =$1.24. 
However, taking into account that for several sources spectral and photometric uncertainties are 
typically of the order of 10\% we adopted as discriminating the slightly higher colour $u - r =$1.4. 
Of course, this simple criterium fails for redshifts higher than $\sim$0.5, and for such objects 
another colour threshold must be adopted.
This problem does not affect largely the analysis of our samples because high $z$ sources are 
very rare among known BL Lacs and are generally so faint in the optical that good photometry 
is not available.

The number of G objects (or with $u - r >$ 1.4) is much smaller than those of N type: they are 
120 (corresponding to about 18\%) in the B0 sample, and 96 in the B1-X sample 
 ($\sim$23\%).

\section{BL Lac types and high energy emission} 

\subsection{The $\Phi_{XR}$ vs $u -r$ plot} 

The bivariate distribution of the B1-X sample in the plane $\Phi_{XR}$ vs $u -r$ 
(Fig. 5), is a very useful tool for improving the classification of BL Lac 
objects.
According to the previous established criteria, the horizontal line, corresponding
to $ Log \Phi_{XR} = -0.5$, discriminates H and L objects, while the vertical one 
having $u -r = 1.4$ discriminates N and G types.
Thus, the resulting four quadrants correspond to HN, HG, LG and LN types, as 
indicated in Fig. 5.
It is well apparent that the large majority of known BL Lac objects are of HN type,
there are also several sources of LN types, while the LG ones are very few.
This is not due to the fact that the X-ray flux is unknown for about 1/3 of the 
sources in the B0 sample, because many of them have low $u -r$ values.
The distribution of points is remarkably not uniform with a high concentration 
in a rather narrow belt crossing the HN and HG regions. 
In the two G regions points are much more sparse with only a relatively denser 
group, centered about at $u - r =$1.95 and Log$\Phi_{XR} =$0.25. 
Ten of the 15 sources in this region are in galaxy clusters or small groups, and, 
if they were excluded because their X-ray emission could not entirely be originated 
in the nucleus, the data concentration would disappear.  


We can define in this plot an approximate region inside which sources' data points are more 
concentrated than outside.
It is rougly limited by the two dashed lines in Fig. 5.
The left $u - r$ boundary of this region appears to be intrinsic to the BL Lac population
and does not depend on selection effects, because there is no reason for excluding sources 
with a high $u$ excess in BL Lac searches in other frequency bands. 
There are, in fact, only very few object on the left side of this belt and they can 
represent the  minimum observed $u - r$ value that is around 0.2 corresponding to a 
spectral index of 0.32.
Note also that the mean $u - r$ colour of HN objects in this region is smaller than for 
LN ones.
This finding can be naturally explained by the fact that the X-ray luminosity is expected 
to be positively correlated with that in the UV range if they are originated by the same
synchrotron component.

In Fig. 5, we introduced some special symbols for different classes of BL Lacs:
red squares mark those sources which are close to or belong to a known cluster of galaxies, 
as derived from NED information or from the recent list based on SDSS images (Hao et al. 
2010), blue circles correspond to objects that in the SDSS images appear, in our opinion, 
to be in small groups, or in close pair or systems, although not reported in the 
literature.
In particular, we searched for other galaxies having an angular separation from 
the BL Lac lower than a few galaxian radii that for a typical elliptical at $z$ about 
0.2 corresponds to 20$''$.
When four or more galaxies were found we classified the source as belonging to a group,
while it was considered a pair (or a small system) when less than three galaxies were in 
the surroundings.
In some cases, redshifts of near galaxies are also available in SDSS and we were able
to verify their actual closeness.

The possibility of confusion of X-ray emission from BL Lacs and galaxy clusters was considered
since early systematic searches as the Einstein Medium Sensitivity Survey (Stocke et al. 
1985, 1991).
The occurrence of BL Lac objects in cluster of galaxies was already investigated by several
authors (e.g. Laurent-Muehleisen et al. 1993, Pesce, Falomo \& Treves 1995; Hart, Stocke 
\& Hallman 2009, Giles et al. 2011) and in some cases the membership was spectroscopically 
confirmed.
The majority of BL Lac objects, however, as found in our sample, appear rather isolated and 
therefore the chance occurrence of a BL Lac in a cluster is not expected to be high, in 
agreement with the previous results of Owen, Ledlow \& Keel (1996), who searched for BL Lac 
objects in a rather large sample of Abell clusters.

The X-ray flux of candidate BL Lacs in clusters can be therefore contaminated by an emission of non-nuclear origin,
and this could lead to the spurious identification of a galaxy with a faint radio emission
as a BL Lac object.
Note that the distribution of these sources in Fig. 5 is highly not uniform.
If the chance to find a BL Lac object in a cluster were independent on its physical
parameters, the number of circles and squares should be higher in the region of high
density of points, while we observe the opposite distribution.
Fig. 6 gives a very clear picture of this effect.
We plotted, for $u - r$ intervals equal to 0.5, the fraction of sources found in
systems (circles and squares) and verified that it is very well correlated with
the red colour of the object.
Objects with $u - r > 3.0$ are all associated with clusters, while the fraction of
associated sources with this coulour index higher than 2.0 is about 0.5.

A list of G-type BL objects having $u - r >$1.8 and including notes about possible
association with clusters is given in Table 1.

\begin{figure}
\vspace{0.0 cm}
\centering
\includegraphics[height=8.4cm,angle=-90]{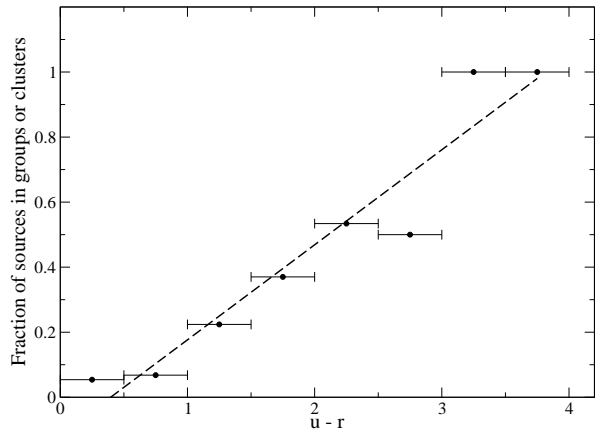}
\caption[]{The fraction of sources in groups or clusters plotted as a function of the colour
$u-r$.
The dashed line is a linear best fit that gives a very good description of the data
(the linear correlation coefficient is 0.96).}
\end{figure}

\subsection{High energy emission of G-type BL Lac objects} 
Bl Lacs are known to be $\gamma$-ray sources.
A significant fraction of them has been detected in the first year of observation 
of Fermi-LAT (Large Area Telescope) and it is reported in the 1FGL catalogue 
(Abdo et al. 2010a).
This number is increased in the very recently released 2FGL catalogue 
\footnote{http://fermi.gsfc.nasa.gov/ssc/data/access/lat/2yr\_catalog/}
In B0 sample there are 126 objects likely associated with 2FGL sources, while 94 are 
in the 1FGL catalogue, nine of which not confirmed in the two-year $\gamma$-ray sky.
All these BL Lacs are marked by a cross in Fig. 5.  
Again, their distribution is far from being uniform: the majority of them is of N-type
and are inside the high density strip.
Note that there are only few LN sources not detected at $\gamma$-ray energies, while 
this does not occur for HN objects; moreover, no possible $\gamma$-ray counterpart is 
reported for LG-type BL Lacs. 
Two examples of SEDs of N sources, detected at high energies, are given in Fig. 2. 

Eight BL Lacs with $u -r > 1.8$ can be associated with $\gamma$-ray sources,
and only the following four sources have $u -r > 2.0$:
BZB~J0319+1845 (2FGLJ0319.6+1849), BZB~J1136+2550 (2FGLJ1137.0+2553), 
BZB~J1532+3016 (1FGL J1531.8+3018), BZB~J2322+3436 (2FGLJ2322.6+3435).
No source with $u -r > 2.25$ is reported to have a possible counterpart at high energies.
One of the sources with $u -r > 2$ (BZB J1532+3016) is a low redshift elliptical galaxy 
($z$=0.065) with a flat spectrum radio emission, that appears likely extended in NVSS map. 
SDSS image shows a companion spiral suggesting an interacting pair.
Considering that the significance of its 1FGL $\gamma$-ray counterpart is equal only to 4.3 
$\sigma$ and that this was not confirmed in the 2FGL, the occurrence of high energy emission
from this G-type object could not be safely established.

The other three sources have redshifts between about 0.1 and 0.2 and appear as isolated elliptical.
Their radio emission, rather faint for two of them, has flat spectra and the $\Phi_{XR}$
values are typical of H-type sources. 
They could then be low luminosity BL Lac nuclei, however a deeper multifrequency study to clarify their nature would be very useful. 

\begin{figure*}
\vspace{0.0 cm}
\centering
\includegraphics[height=8.4cm,angle=-90]{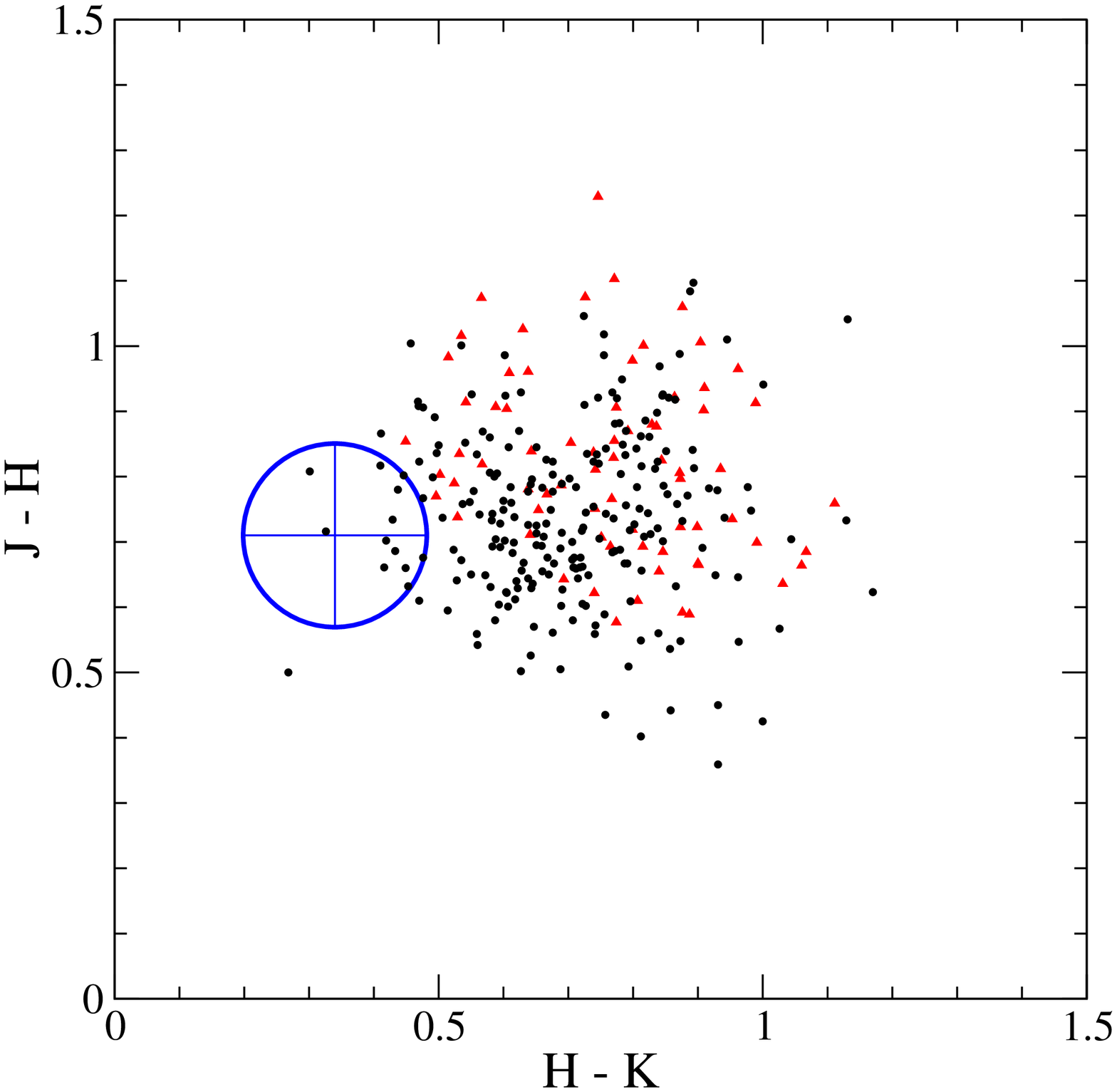}
\includegraphics[height=8.4cm,angle=-90]{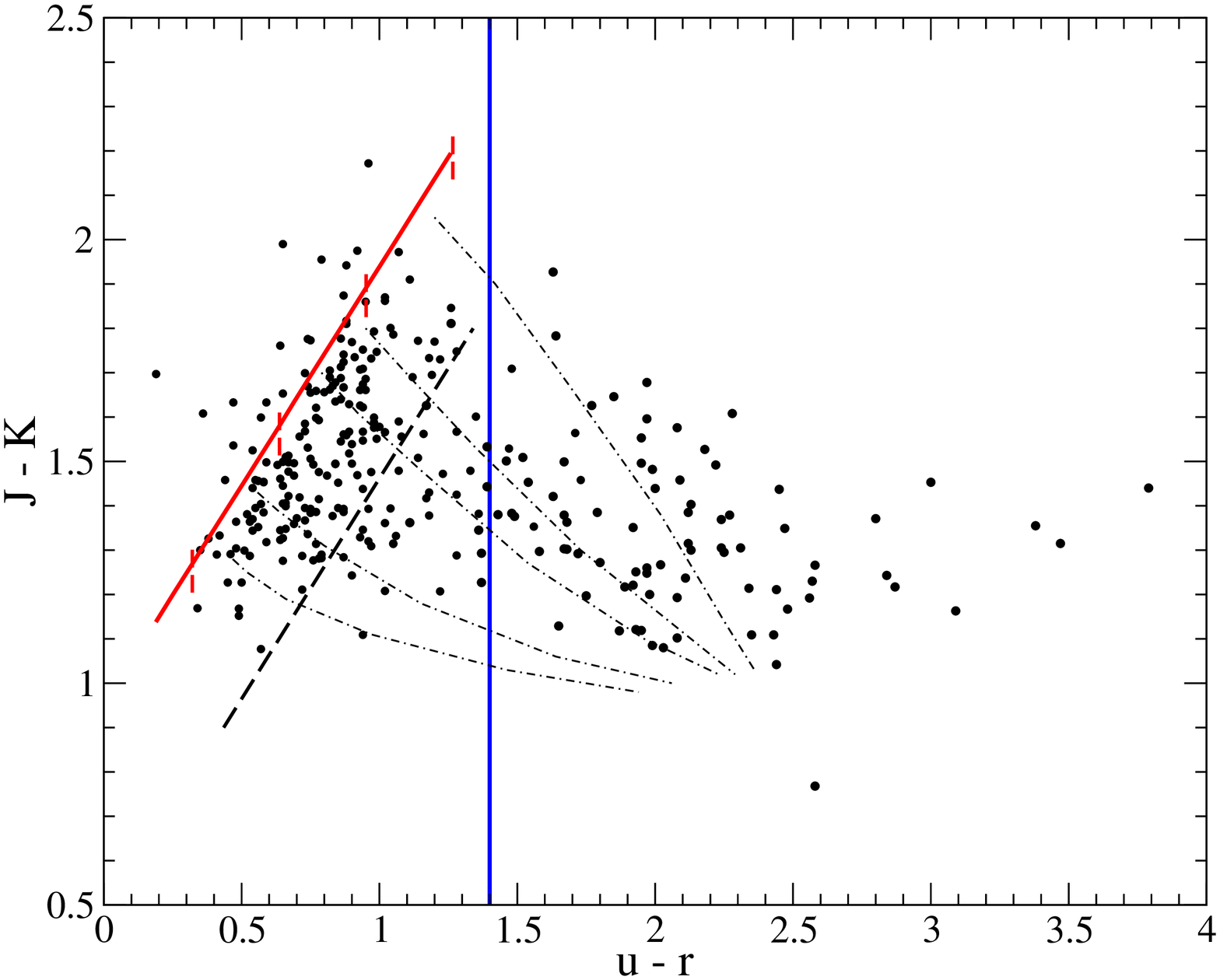}
\caption[]{Left panel: the J$-$H, H$-$K colour plot of BL Lac objects, the large open circle
is a region of radius equal to 1 standard deviation of the population and where the majority
of normal elliptical galaxies from the Uppsala catalogue clusterize; filled black circles are
sources with a reported X-ray detection, while red triangles indicate objects not yet detected.
Right panel: plot of the infrared J$-$K colour vs $u - r$; the thick red line is the locus of
power law energy spectra and the vertical short lines mark spectral index values of 2.0, 1.5,
 1.0, 0.5 (top to bottom); the blue line corresponds to $u - r=$1.4 and discriminates N and G
type sources.
Thin dot-dashed lines describe the change of colours when a synchrotron component is added
to a normal elliptical galaxy with a variable brightness ratio.
Top to bottom the peak frequencies of the non-thermal component vary from Log$\nu_p$=12.5 to
15.5, with a parabolic curvature coefficient $b$ equal to 0.1 or 0.2.}

\end{figure*}

\section{Infrared colours}

Another frequency band in which homogenueous data sets are available for a large
number of BL Lac objects is the near infrared.
We investigated the emission properties of our sample in this band to extend the
available information useful for identifying and properly classifying these sources. 
We searched for the B0 sources in the 2MASS point source catalogue, 
imposing a positional coincidence better than 3 arcsec. 
Given the limiting magnitude of the 2MASS ($J \simeq$16) only the brightest sources 
(319) are present in this catalogue.
A further selection was made on the photometric quality, which is flagged with a 
letter, in decreasing quality order: only sources with quality flag A 
or B in all the $J$, $H$, $K$ bands were considered to select the subsample B1-2M 
containing 281 objects. 

Chen, Fu \& Gao (2005) studied the colour distribution of a rather large sample 
of BL Lac objects and found that, in the ($J-H$, $H-K$) diagram, they are 
simmetrically clustered around a point at the position (0.7; 0.7). 
Our analysis shows that B1-2M sources have a similar pattern around the same 
centroid (Fig. 7).
We compared this distribution with that of normal elliptical galaxies using $J$, $H$, 
$K$ magnitudes of a sample of rather nearby galaxies selected from the Uppsala 
Galaxy Catalogue (UGC, Nilson 1973) as reported in the 2MASX extended sources catalogue. 
Their colour distribution is centered approximately at $J-H$=0.7, $H-K$=0.3 and 
exhibits a nearly circular pattern with a standard deviation of 0.14 magnitudes 
on both colours (the large circle in the left panel of Fig. 7).
Rather surprising, there is no well apparent difference between the distributions
of N and G-type BL Lacs, being the $H-K$ colour of latter ones redder than
normal elliptical galaxies.
In principle, one could expect some difference in the colours of these two types 
because these classes are based on the relative brightness of the optical nuclear 
emission to the host.
Indeed, the nuclear component strength is based on the level of the continuum at 
optical wavelengths shorter than Ca H\&K break feature, so that G-type sources 
are expected not to exhibit a high non-thermal component also in the near Infrared 
bands. 

We explored if this result could be explained by the effect of a k-correction, due to 
the source redshifts, on the colour indices of our BL Lac objects.
This correction is not relevant if the spectrum is dominated by the non-thermal emission 
(e,g, it is locally described by a power law), but it may be a concern for a substantial 
stellar contribution, i.e. for objects classified as G-type. 
A suitable $k$-correction for elliptical galaxies in the $J$, $H$, $K$ magnitudes 
is reported by Poggianti (1997): up to redshift $z=$0.3 the effect on the $J-H$ colour 
is less than 0.05 magnitudes, while the $H-K$ colour becomes redder with increasing $z$, 
by up to 0.15 for $z$=0.3.
This correction is anyway always smaller than 0.2 in both colours for $z$ less than 0.6.
Of our 78 G-type sources with good quality (A or B) 2MASS photometry, one has no 
firmly established redshift and no $k$-correction can be computed, and another one only 
has $z\ge0.3$. 
We are therefore forced to think that the redder $H-K$ colours of the G-Type blazars 
can only partially be explained by their redshifts but it is largely intrinsic. 

The right panel of Fig. 7 shows the $J-K$ colour plotted against $u-r$.
As expected from the previous result, the mean value of the former colour is around 
1.5 and there is an evident concentration of points inside the strip marked by a thick 
solid red line, representing the colours of a single power law over the near IR to $u$ 
band, and a parallel dashed line.
According to our simple criterium, all BL Lacs inside the strip have $u-r < 1.4$ and 
are of N-type.
Note that the density of data points is high close to this line and that only a small 
number of sources has bluer colours.
The fact that sources tend to be redder than a simple power law indicates that their 
spectral distributions in the optical-UV are steeper than in the near infrared, 
as expected if they are described by a parabolic law in the log scales, in agreement 
with the early findings by Landau et al. (1986) and by more recent works (see,
for instance, Massaro et al. 2004a, 2004b, 2006, Tramacere et al. 2011).

To reproduce the observed distribution in the $u-r$, $J-K$ plane we considered 
a SED given by the sum of an average elliptical galaxy (Buzzoni 2005) and a non-thermal 
component described by a log-parabola $ Log(\nu F_{\nu}) = A - b [Log(\nu/\nu_p)]^2 $.
Typical values for $b$ are in the range 0.1-0.3, while the peak frequency $\nu_p$ is 
higher than $\sim$10$^ {13}$ Hz for L BL Lac objects (Maselli et al. 2010b) and may
be even higher than $\sim$10$^ {16}$ Hz for H sources.
The results of our simulations are also reported in Fig.7: each of the dashed-dotted 
lines corresponds to a couple of values of the parameters Log$(\nu_p)$ (12.5, 13.5, 15.5) 
and $b$ (0.1, 0.2), and the colour change is due to variation of the log of the 
ratio between the galaxy to non-thermal flux at the $K$ band from $-$4 to 1.
With decreasing non-thermal contribution the object colour moves out of the power-law 
line to reach the point corresponding to a normal elliptical.
The fact that all curves converge towards the point (1.0; 2.4) clearly depends on the 
adopted elliptical galaxy model; a model with a redder $u-r$ would move the curves on
the right side of the plot thus including the data points of several G-type BL Lacs.
Moreover,a local reddening in the host galaxy can also increase the $u-r$ values.
We recall that a diffrence between all the considered sources and normal elliptical
galaxies is that the former ones have a nuclear radio emission, that could be
related to the $K$ band excess: for instance, when the synchrotron component peaks
at a rather low frequency and is sufficiently curved. 
%

\section{Discussion}

The most important result of our analysis is the assessment of more precise 
observational criteria to define the properties of BL Lac objects.
In addition to the known L and H classes, we introduced the N and G classes 
to take into account the relative power of the nuclear emission to host galaxy.
With the general accepted statement that a blazar, and more specifically a BL Lac, 
is an active nucleus of a galaxy whose emission is dominated by non-thermal processes 
and boosted by relativistic motions, it is important to know if the nucleus is or not
bright enough to overcome the host galaxy contribution, at least in frequency ranges 
where the latter is typically low, in particular, at frequencies higher than the 
Ca H\&K break.
Our analysis showed that a simple estimate of the dilution of the Ca H\&K break 
contrast cannot always be considered a good tool for searching and properly 
classifying this type of AGNs.
In fact, the contrast is based on measures of the source flux in rather small 
wavelength intervals on the two sides of the break, and does not takes properly into 
account if the nuclear emission turns out or not to be the dominant one for 
increasing frequencies.
Under this respect, the colour $u - r$ is a better indicator of the N or G type, 
at least for sources with a redshift $z < 0.5$.
On this basis we can conclude that a fraction of sources, reported in the literaure
as genuine BL Lacs, could not belong to this class of AGNs and more accurate investigations
of their individual properties and enviroment must be performed to safely establish 
whether they are genuine BL Lac objects.
The 77 G-type BL Lac objects reported in Table 1 were discovered in many surveys, 
confirming that these rather broad selection criteria are largely adopted and that the 
possibility of considering galaxies without a clear evidence of relativistic beaming as 
genuine BL Lac objects is not negligible.
Sources occasionally reported in the literature as ``weakly beamed'' BL Lacs should not 
be properly considered blazars. 

The presence of a faint radio emission could be due to an origin different from a
beamed relativistic jet and, therefore, cannot be considered a definitive test.
Several other types of AGNs, e.g. radio galaxies, Seyfert galaxies, etc...
are radio emitting with a large spread of power, but they do not exhibit blazar 
properties.
As for concerning multifrequency studies, our analysis demonstrates that the 
detection of an X-ray flux from the direction of the candidate blazar can be 
confusing if the source environment is unknown.
The uncompleteness of catalogues of galaxy clusters, particularly for the 
small ones, make difficult to establish if the X-ray emission has been incorrectly 
associated with the nuclear activity leading to wrong BL Lac identifications.
Another source of confusion can be due to the radio emission observed from
some clusters of galaxies (Giovannini et al. 2009).
This emission is generally faint and with a steep spectrum, but when BL Lac association 
is derived from correlations between catalogues at a single frequency and with a limited 
angular resolution, like NVSS (Condon et al. 1998), spurious results are possible.

Values of the optical and infrared colours $J - K > 1.1$ and $u - r < 1.4$ can provide 
a simple test to obtain a first indication that an AGN has a non-thermal continuum 
overwelming the host galaxy emission, hence it is a possible blazar, or a BL Lac if 
emission lines are absent.
The fact that the majority of sources with a $\gamma$-ray counterpart has colours within 
the two rather narrow strips in Fig. 5 and Fig. 7, supports the conclusion that sources 
whose colours are well outside these regions cannot be safely considered BL Lacs. 
Values beyond these limits do not exclude the blazar nature of the object, but they 
must be considered as a clear indication that further investigations are necessary.

The use of these two photometric colours can be of practical relevance for two problems.
The first problem concern the existence of ``radio faint'' (or ``radio silent'') BL Lacs.
A rather large fraction of known BL Lac objects, particularly HBL ones, has a radio 
flux at 1.4 GHz lower than 20 mJy or even 10 mJy; moreover, they are generally 
undetected in the major 5 GHz surveys.
We recall that the recent BL Lac searches by Plotkin et al. (2008, 2010) report several 
candidates with radio fluxes lower than 5 mJy, which were not included in the Roma-BZCAT 
(3rd edition).
The 2-colour test proposed here can be then a further selection tool for good 
candidates.
Note than hot stars as white dwarfs are easily excluded, in addition to their 
appreciable proper motion, because their $J - K$ is much lower than that of BL Lacs.
Of course, a simple colour-based method is not able to discriminate BL Lac objects 
from quasars and optical spectroscopy is necessary to search for the occurrence of 
emission lines and a proper classification of the source.
In a very recent analysis, Massaro et al. (2011b) found that a large fraction of 
BL Lac objects are located in a rather narrow strip in the [3.4]--[4.6]--[12] $\mu$m 
two colour diagram derived from magnitudes measured by the {\it Wide Infrared Survey 
Explorer (WISE)} satellite (Wright et al. 2010). 
This strip is particularly well marked by HBL objects and many of them are of N type.
We expect, therefore, that the use of mid-infrared photometry will be a very promising and
powerful tool to distinguish N and G sources, to gain more information on the non-thermal 
nuclear emission and the distributions in the $J - H$, $H - K$ plane.

The second problem is the search for possible counterparts in the error boxes of 
$\gamma$-ray sources. 
In the LAT catalogues (1FGL, Abdo et al. 2010a, and 2FGL, Abdo et al. 2011) there are 
a few hundredths sources without a reasonable counterpart to associate with.
The finding of a source having colours satisfying the above limits, possibly associated 
with a radio or X-ray emission, would be useful for selecting interesting candidates
as in the case considered by Maselli et al. (2011).
Similar research could be applied to the study of serendipitous sources detected 
inside the fields of X-ray imaging telescopes, like Chandra and Swift-XRT.   
We stress that the colour based test is not a new approach.
It is, in fact, an updated and revised version of the method proposed and successfully 
used by Braccesi (1967) and Braccesi et al. (1968) in the sixsties for searching 
quasars, and particularly the radio-quiet ones.
The existence of large and easily accessible databases, makes the colour test 
particularly simple and powerful.

\section*{Acknowledgments}
We are grateful to Gino Tosti and Andrea Tramacere for useful
discussions.
We acknowledge the financial support 
from Universit\`a di Roma La Sapienza. 
In this work we used the online version of the Roma-BZCAT
and of the scientific tools developed at the ASI Science
Data Center with the supervision of Paolo Giommi.
This publication made use of the Sloan Sky Digital Survey, the
NED database and of other astronomical catalogues distributed in 
digital form at CDS Strasbourg.

\newpage

\begin{table*}
\caption{List of G-type BL Lac objects in the B0 sample having $u - r >$ 1.8. }
\label{tab:GBL1}
\begin{tabular}{llrrrrll}
\hline
Source & $u-r~~$ & $J-K$ & $S_{1.4}$ & $Log~ \Phi_{XR}$ & $z$~~ & Ref & Notes \\
       & mag   & mag   & mJy       &                  &       &     &      \\
\hline
 BZB J0001$-$1031  &  2.3 & 1.30 & 41  &           & 0.252 & PL8 &  \\ 
 BZB J0056$-$0936  &  1.9  & 1.22 & 199 &     0.324 & 0.103 & RBS & gr \\
 BZB J0103+1526    &  2.0  & 1.60 & 225 &           & 0.246 & CL5 &  \\
 BZB J0110+4149    &  2.4  & 1.04 & 85  &     0.139 & 0.096 & RAX & gr \\
 BZB J0148+1402    &  2.9  & 1.22 & 44  &  $-$0.125 & 0.125 & PG5 & cl: GMBCG \\
 BZB J0232+2017    &  1.8  & 1.27 & 82  &     0.837 & 0.139 & PG5 & gr \\
 BZB J0319+1845    &  2.1  & 1.24 & 21  &     1.598 & 0.190 & PG5 & F2 \\  
 BZB J0737+3517    &  3.0  & 1.45 & 26  &     0.610 & 0.213 & RAX & gr \\
 BZB J0741+3205    &  2.0  & 1.48 & 77  &           & 0.179 & PL8 & cl: GMBCG \\
 BZB J0745+3312    &  3.8  & 1.44 & 22  &  $-$0.755 & 0.220 & ROX & cl: MaxBCG, GMBCG \\
 BZB J0751+2913    &  2.8  & 1.37 & 10  &     0.322 & 0.185 & ROX &  \\
 BZB J0754+3910    &  1.9  & 1.22 & 58  &  $-$0.120 & 0.096 & BR0 & ss \\
 BZB J0758+2705    &  2.0  & 1.27 & 54  &           & 0.099 & PL8 & ss \\
 BZB J0809+3455    &  2.0  & 1.08 & 223 &     0.261 & 0.083 & CLS & F1, cl: GMBCG  \\
 BZB J0810+2846    &  2.0  &      & 42  &           & 0.272 & PL8 &  \\
 BZB J0810+4911    &  3.1  & 1.16 & 11  &     0.260 & 0.115 & CLS & cl: MaxBCG, GMBCG \\
 BZB J0823+1524    &  2.1  & 1.40 & 22  &           & 0.167 & PL0 &  \\
 BZB J0828+4153    &  2.0  & 1.64 & 91  &     0.094 & 0.226 & RAX & cl: GMBCG \\
 BZB J0850+3455    &  1.9  & 1.25 & 35  &     0.269 & 0.145 & RAX & F1 \\
 BZB J0856+5418    &  4.3b & 1.28 & 57  &     0.338 & 0.259 & RAX & cl:  MACS, GMBCG \\
 BZB J0903+4055    &  1.9  & 1.12 & 36  &     0.745 & 0.188 & RAX &  \\
 BZB J0927+5327    &  3.1  & 0.93 & 5   &     1.334 & 0.201 & RAX & cl: GMBCG \\
 BZB J0927+5545    &  2.5  & 1.44 & 78  &  $-$0.637 & 0.221 & ROX & ss \\
 BZB J0932+3630    &  2.5  & 1.17 & 52  &  $-$0.136 & 0.154 & PL8 & ss \\
 BZB J1001+2048    &  2.5  & 1.36 & 2   &     0.929 & 0.344 & EMS & gr? \\
 BZB J1012+3932    &  2.0  & 1.26 & 20  &     0.512 & 0.171 & ROX &  \\
 BZB J1018+3128    &  2.4  & 1.21 & 4   &     1.322 & 0.161 & RAX &  \\
 BZB J1022+5124    &  2.3  & 1.61 & 5   &     1.838 & 0.142 & EMS &  \\
 BZB J1028+0555    &  2.0  & 1.84 & 19  &     0.022 & 0.234 & PL8 &  \\
 BZB J1028+1702    &  2.2  & 1.30 & 79  &  $-$0.053 & 0.169 & SE5 &  \\
 BZB J1033+4222    &  2.0  & 1.22 & 45  &  $-$0.222 & 0.211 & CLS & cl: GMBCG \\
 BZB J1041+1324    &  1.8  & 1.64 & 57  &           & 0.375 & PL8 & cl: GMBCG \\
 BZB J1041+3901    &  2.1  & 1.10 & 33  &     0.385 & 0.210 & PL8 & cp \\
 BZB J1053+4929    &  2.0  & 1.12 & 64  &     0.108 & 0.140 & RAX & F2, cl: MS, GMBCG \\
 BZB J1059+4343    &  2.0  & 1.77 & 40  &  $-$0.301 & 0.459 & PL8 &  \\
 BZB J1124+5133    &  2.6  & 1.27 & 54  &           & 0.235 & CLS & gr \\
 BZB J1136+2550    &  2.3  & 1.29 & 16  &     0.468 & 0.156 & RAX & F2 \\
 BZB J1145$-$0340  &  2.1  & 1.30 & 18  &     1.358 & 0.167 & SH5 &  \\
 BZB J1156+4238    &  2.6  & 1.19 & 14  &     0.693 & 0.172 & EMS & cl: MS \\
 BZB J1157+2822    &  1.9  & 1.65 & 31  &     0.341 & 0.300 & PL0 &  \\
 BZB J1201$-$0007  &  2.2  & 1.37 & 70  &     0.176 & 0.165 & RAX & cl: [EAD]238, GMBCG\\
 BZB J1201$-$0011  &  2.4  & 1.11 & 28  &     0.301 & 0.164 & ROX & cl:  MaxBCG, GMBCG \\
 BZB J1203+6031    &  2.0  & 1.25 & 190 &  $-$0.404 & 0.065 & CLS & F2 \\
 BZB J1221+4742    &  2.1  & 1.19 & 42  &     0.364 & 0.210 & RAX & ss \\
 BZB J1223+4650    &  2.1  & 1.58 & 13  &     0.442 & 0.261 & ROX & cl: NSCS, GMBCG \\
 BZB J1238+5406    &  2.0  & 1.55 & 39  &           & 0.224 & PL8 &  \\
 BZB J1243+5212    &  2.5  & 1.35 & 38  &  $-$0.404 & 0.200 & ROX &  \\
 BZB J1243$-$0613  &  2.7  & 1.22 & 345 &           & 0.139 & PKQ & ss gr?\\
 BZB J1253+0326    &  1.9  & 1.12 & 107 &     0.193 & 0.066 & RAX &  \\
 BZB J1331$-$0022  &  1.9  & 1.74 & 9   &     0.523 & 0.243 & CL5 &  \\
 BZB J1348+0756    &  2.0  & 1.50 & 53  &     0.400 & 0.250 & PL0 & cl: GMBCG \\
\hline
\end{tabular}
\end{table*}


\setcounter{table}{0}
\begin{table*}
\caption{G type BL Lac objects having $u - r >$ 1.8 (continued)}
\label{tab:GBL2}
\begin{tabular}{llrrrrll}
\hline
Source & $u-r~~$ & $J-K$ & $S_{1.4}$ & $Log~ \Phi_{XR}$ & $z$~~ & Ref & Notes \\
       & mag   & mag   & mJy       &                  &       &     &       \\
\hline
 BZB J1401+1350    &  3.5  & 1.32 & 8   &     0.787 & 0.212 & ROX & cl: MaxBCG, GMBCG \\ 
 BZB J1424+3705    &  2.2  & 1.53 & 79  &  $-$0.244 & 0.290 & RAX & cs \\
 BZB J1427+5409    &  2.6  & 0.77 & 44  &  $-$0.064 & 0.106 & RAX & gr \\
 BZB J1427+3908    &  2.0  & 1.11 & 7   &     0.359 & 0.165 & PL8 &  \\
 BZB J1435$-$0055  &  1.9  & 1.35 & 15  &           & 0.285 & PL8 &  \\
 BZB J1436+4129    &  2.8b &      & 53  &           & 0.404 & PL8 &  \\
 BZB J1445+0039    &  3.7b & 1.68 & 8   &     0.903 & 0.306 & ROX & cl: CE J221.312271+00.651836 \\
 BZB J1502+2528    &  2.2  &      & 51  &           & 0.178 &     &  \\
 BZB J1510+3335    &  2.6  & 1.23 & 8   &     1.500 & 0.114 & RAX &  \\  
 BZB J1516+2918    &  2.3  & 1.21 & 134 &  $-$0.023 & 0.130 & RAX & gr \\
 BZB J1532+3016    &  2.0  & 1.08 & 52  &     0.872 & 0.065 & RAX & F1, ss \\
 BZB J1544+5017    &  1.9  &      & 29  &  $-$0.617 & 0.494 & ROX &  \\ 
 BZB J1604+3345    &  2.8  & 1.24 & 7   &     1.012 & 0.177 & RAX &  \\
 BZB J1616+3756    &  1.9  & 1.38 & 4   &     1.357 & 0.202 & SH5 &  \\
 BZB J1624+3726    &  2.2  & 1.38 & 59  &  $-$0.324 & 0.199 & RAX &  \\ 
 BZB J1628+2527    &  2.0  & 1.44 & 72  &     0.292 & 0.220 & SE5 & cl: GMBCG \\
 BZB J1637+4547    &  2.0  & 1.68 & 45  &     0.166 & 0.192 & CL5 & cl: MaxBCG \\ 
 BZB J1643+2131    &  3.6  & 1.30 & 8   &     0.837 & 0.154 & ROX & cl: GMBCG \\
 BZB J1647+2909    &  2.1  & 1.31 & 390 &  $-$0.958 & 0.132 & MB6 & ss \\
 BZB J1655+3723    &  1.9  &      & 76  &           &       & PL0 &  \\
 BZB J1717+2931    &  3.1  & 1.43 & 3   &     2.040 & 0.276 & ROX & cl: RBS 1634 \\ 
 BZB J1750+4700    &  2.2  & 1.49 & 99  &     0.312 & 0.160 & LM9 & cl: MACS \\
 BZB J2145$-$0434  &  2.4  & 1.11 & 67  &     0.464 & 0.070 & RAX &  \\
 BZB J2248$-$0036  &  2.1  & 1.46 & 61  &           & 0.212 & CL5 & cl: GMBCG \\
 BZB J2322+3436    &  2.1  & 1.38 & 95  &     0.060 & 0.098 & LM9 & F2 \\
 BZB J2350+3622    &  2.0b & 1.68 & 317 &  $-$1.422 & 0.317 & DXR &  \\
\hline
\end{tabular} \\

Notes: \\
b: value with a large uncertainty; \\
cl: cluster of galaxies; \\
cp: close pair of galaxies; \\
gr: group of galaxies; \\
F1: possible counterpart of a Fermi-LAT 1FGL source \\
F2: possible counterpart of a Fermi-LAT 2FGL source \\
ss: small system of close galaxies. \\

References: \\
CL5 = Collinge et al. (2005) - SDSS; \\
CLS = March\~a et al. (2001); Caccianiga et al. (2002) - CLASS; \\
DXR = Perman et al. (1998); DXRBS \\
EMS = Stocke et al. (1991), Rector et al. (1999) - EMSS; \\
LM9 = Laurent-Muehleisen et al. (1999) - RGB;  \\
MB6 = March\~a et al. (1996); \\
PG5 = Padovani \& Giommi (1995);\\
PKQ = Jackson et al (2002), Hook et al (2003) - Parkes Quarter Jansky \\
PL0 = Plotkin et al. (2010) - SDSS;  \\
PL8 = Plotkin et al. (2008) - SDSS; \\
RAX = Kock et al. (1996), Wei et al. (1999), Bade et al. (1998), Voges et al. (1999), Bauer et al. (2000), Brinkmann et al. (2000) - RASS, ROSAT; \\
ROX = Turriziani et al. (2007) - ROXA, http://www.asdc.asi.it/roxa/; \\
SE5 = Sowards-Emmerd et al. (2005); \\
SH5 = Giommi et al. (2005), Piranomonte et al. (2007) - Sedentary Survey \\

\end{table*}

\noindent

\label{lastpage}
\end{document}